\documentclass[fleqn,usenatbib]{mnras}

\usepackage{newtxtext,newtxmath}
\usepackage{orcidlink}
\usepackage[T1]{fontenc}

\DeclareRobustCommand{\VAN}[3]{#2}
\let\VANthebibliography\thebibliography
\def\thebibliography{\DeclareRobustCommand{\VAN}[3]{##3}\VANthebibliography}

\usepackage{graphicx}
\usepackage{amsmath}	
\usepackage{float}
\usepackage{tabularx}
\usepackage{multirow}
\usepackage{verbatim}
\usepackage{tikz}
\usetikzlibrary{arrows.meta, positioning, shapes.geometric}
\usepackage{lineno}

\newcommand{\axaf}{\mbox{\em Chandra\/}}

\newcommand{\gaia}{\mbox{\em Gaia\/}}

\newcommand{\hst}{{\it HST}}

\title[Speeding up Gravitational Lens Mass Models with Machine Learning: Applications in X-ray Astronomy]{Speeding up Gravitational Lens Mass Models with Machine Learning: Applications in X-ray Astronomy}

\author[A. Ostridge et al.]{
Alex Ostridge\orcidlink{0009-0008-1273-9562}$^{1,2}$\thanks{Corresponding author: alex.ostridge@outlook.com},
Rafael Mart\'{i}nez-Galarza\orcidlink{0000-0002-5069-0324}$^{1}$,
J\'{u}lia M. Sisk-Reyn\'{e}s\orcidlink{0000-0003-3814-6796}$^{1}$\thanks{Corresponding author: julia.sisk\_reynes@cfa.harvard.edu},
Daniel A. Schwartz\orcidlink{0000-0001-8252-4753}$^{1}$ and \newauthor
Anna Barnacka\orcidlink{0000-0001-5655-4158}$^{1}$\\
$^{1}$Center for Astrophysics Harvard \& Smithsonian, 60 Garden St, Cambridge, MA 02138\\
$^{2}$University of Southampton, University Road, Southampton, SO17 1BJ, UK}

\begin{document}
\label{firstpage}
\pagerange{\pageref{firstpage}--\pageref{lastpage}}
\maketitle

\begin{abstract}
Multi-wavelength observations of quadruply lensed quasars constitute a powerful probe of cosmology, dark matter substructure along the line of sight, and the structure of X-ray emitting regions in high-redshift quasars. These investigations are conditional on acquiring an accurate model for the surface mass density of matter lensing these quasars. We propose a simulation-based machine learning method to accelerate parameter inference in real quadruply lensed systems by several orders of magnitude. We simulate a grid of quadruply lensed sources with Singular Isothermal Ellipsoid (SIE) lenses and use the projected positions of the four lensed images to train two fully connected neural networks that predict the mass parameter and ellipticity. For a large fraction of simulated systems, the neural network-initialised mass models converge in time-scales of a few minutes and recover the source position at the $<0\farcs005$ level for a broad range of lens masses and ellipticities. We apply our neural networks to seven quadruply lensed quasars, lensed by isolated galaxies or a galaxy-perturber pair, which have archival \axaf\ observations. The final optimised mass models for each quasar predict the observed lensed image positions in \gaia\ Data Release 3. These mass models enable the caustic method, which locates the X-ray-to-optical emission regions to milliarcsecond precision in these otherwise unresolvable systems, improving the effective angular resolution of \axaf\ at high-$z$ by up to two orders of magnitude. Our approach accelerates this mass modelling by supplying informed initial parameters, enabling application to the many new quadruply lensed systems expected from forthcoming surveys.
\end{abstract}


\section{Introduction} \label{Sec:Introduction}

\subsection{Motivation}
Active galactic nuclei (AGN) are compact regions found at the centres of 1--10 per cent of massive galaxies which achieve extreme luminosities across the electromagnetic spectrum. It is now believed that all supermassive black holes (SMBH) will undergo a phase as AGN. Their X-ray emission traces the innermost accretion flow and the processes driving SMBH--galaxy co-evolution. With its $0\farcs5$ on-axis angular resolution, \axaf\ has resolved relativistic jets and ultra-fast outflows at $z \ll 1$ at parsec to hundred parsecs scales. Such jets are clear signatures of AGN feedback that inject energy and momentum into the host interstellar medium and regulate star formation \citep{2013_kormendy_ho,2012_agn_feedback_review_fabian}.

Offset AGN and dual AGN are observationally elusive, yet predicted outcomes of hierarchical structure formation \citep{volonteri_2016_horizon_agn}. While offset AGN could be signatures of recoiling SMBHs ejected following mergers with associated anisotropic gravitational wave emission \citep{blecha_2016_offset}, dual AGN are indicative of major galaxy mergers. If gravitationally bound, an AGN pair will coalesce and yield gravitational-wave signatures detectable by current or future gravitational-wave observatories \citep{volonteri_2016_horizon_agn,chen_2023_offset_agn}. However, at cosmological distances $> 450 \ \mathrm{Mpc}$ (i.e. $z > 0.1$), sub-kpc scales cannot be directly resolved by any existing -- or even conceived -- X-ray telescope.

\citet{barnacka_galaxies_2017, barnacka_gravitational_2018} demonstrated that the flux magnification and spatial amplification of strong gravitational lensing of AGN (including quasars) can boost the effective spatial resolution of \axaf\ at high redshift by up to two orders of magnitude. These papers introduced the caustic method -- a technique that locates the origin of multiwavelength emission in lensed AGN to milliarcsecond precision at cosmological distances. At $z \approx 1$, 10 milliarcseconds corresponds to a projected distance $\approx 80 \ \mathrm{pc}$. Therefore, the caustic method offers a unique opportunity to search for sub-kpc spatial offsets in AGN at early cosmic epochs around the peak of star formation across a range of spatial scales that remain inaccessible to direct imaging.

\subsection{The caustic method}
Gravitational lensing is the bending of light from a distant object (the `source') caused by the gravitational field of a massive foreground object (the `lens' or `deflector'). Strong lensing is the regime of lensing where the source, lens, and observer are closely aligned, resulting in multiple lensed images of the source. Here, we focus on systems where galaxies containing an AGN are lensed by either isolated galaxies or an isolated galaxy with a nearby perturber. The separation between pairs of lensed images is typically of the order of $\sim 1\farcs0$, and is thus typically only resolvable with instruments with comparable or finer angular resolution. At X-ray wavelengths, \axaf\ is the only instrument able to spatially resolve individual lensed images of such systems. In systems where the source is inside the caustic (loci of source positions where the flux magnification is maximal), four lensed images are typically observed. Whilst the effective potential of the lens determines the resulting lensed image configuration, their respective fluxes may be affected by stars and compact objects in the lensing galaxy. In addition, substructure along the line of sight such as dark matter halos may perturb the observed image positions at the milliarcsecond level \citep{Nierenberg_2014, nierenberg_2017, Chen_2007, Inoue_2012}.
 
We are interested in applying the caustic method to a sample of lensed AGN at redshifts $z = 1$--3 with archival \gaia\ optical and \axaf\ X-ray observations. We seek to measure spatial offsets between the X-ray and optical emission regions across the sample to look for evidence of AGN outflows, radio jets, and AGN pairs \citep[for recent results, see][]{siskreynes_2026,rogers_2025_milliarcsecond, spingola_milliarcsecond_2022, schwartz_resolving_2021}. Pursuing caustic-method lensing of AGN with multiwavelength observations is a promising avenue given that $\approx 10^{3-4}$ new lensed AGN are expected to be revealed by successive \gaia\ data releases and by ongoing and upcoming surveys with \textit{Euclid}, \textit{Rubin}, \textit{Roman}, and in the radio with DSA-2000 and SKA \citep{collett_population_2015,oguri_gravitationally_2010,weiner_predictions_2020,McCarty_2025_dsa-2000,wu_polishing_lenses_2026}. Using the caustic method on multi-wavelength datasets on these new lensed AGN will enable systematic searches for high-redshift analogues of low-$z$ feedback phenomena and reveal unique distant, sub-kpc rare systems -- offset AGN, compact AGN pairs, and other signatures of SMBH activity.

The first step required by the caustic method is to infer a model for the projected surface mass density of the deflector (a `mass model'\footnote{Throughout this paper, we use the term `mass model' to refer to the functional form of the adopted mass distribution, and the term `lens configuration' to refer to a specific realisation of the mass model defined by a particular set of mass parameter and ellipticity values.}) such that the model-predicted quasar lensed image positions match the observed positions within their corresponding astrometric uncertainties. Quadruply lensed AGN -- comprising $\sim 5$--10 per cent of all known lensed AGN -- in \gaia\ are particularly suited to this purpose: the four lensed image positions provide an overdetermined coordinate system to which the X-ray data can be robustly referenced -- doubly lensed systems do not provide sufficient constraints for this purpose -- and \gaia's state-of-the-art milliarcsecond astrometry enables source plane reconstruction to within the same precision. The second step analyses the \axaf\ observations and exploits lensing caustics as non-linear spatial amplifiers \citep{barnacka_galaxies_2017,barnacka_gravitational_2018}. The third step considers a grid of alternate positions for the X-ray source, each of which is forward-modelled through the fixed lensing geometry to predict a distinct set of lensed image positions. Since lensing is achromatic, co-spatial X-ray and optical emission regions must produce identical lensed image configurations for a fixed mass model. The final step addresses which alternate source best reproduces the observed photon distribution (per sky pixel) in the \axaf\ observations by using a double-maximum likelihood approach, subsequently determining the allowed region for the X-ray source relative to the optical with milliarcsecond precision. This yields an effective angular resolution for \axaf\ up to two orders of magnitude finer than its native $\sim0\farcs5$.

\subsection{Machine learning in strong lensing}
Obtaining mass models of galaxy-galaxy lensed systems suitable for the caustic method is time-intensive, often requiring several days per system. For a given parametrisation of the effective gravitational potential of the lens, this runtime is dominated by the search for initial parameters that, once optimised, yield a mass model that reproduces the observed image positions to within astrometric uncertainties. The optimisation is further slowed down by the need to minimise a large number of free parameters under limited observational constraints (i.e. the RA, DEC values of each lensed image). At present, suitable initial parameter values are obtained by combining observational priors with educated guesses, before iteratively optimising until a pragmatic $\chi^{2}$ convergence criterion (that is, the astrometric precision of the survey instrument) is satisfied. Emerging machine learning and inference techniques offer a promising route to significantly compress this timeline by inferring suitable initial conditions of forward mass modelling.

Most current research using ML for gravitational lensing applications focuses on either searching for new lens candidates in mock and real data \citep{schaefer_deep_2018, davies_using_2019,petrillo_testing_2019, canameras_holismokes_2021, andika_streamlined_2023, li_csst_2024}, or optimising parameter inference in known lenses assuming a lens model from the literature \citep{hezaveh_fast_2017, schuldt_holismokes_2021, wagner-carena_hierarchical_2021, huang_strong_2022, biggio_modelling_2023}. Convolutional Neural Networks (CNNs) are well suited for imaging-based lensing works since they use an architecture of convolutional neuron layers that are responsive to a specific sub-field of the image, thus profiting from spatial correlations between the pixels in the image, creating representations that are powerful for classification and regression tasks.

Recent work using CNNs in strong lensing research has centred on identifying and subsequently modelling gravitationally lensed candidates in \textit{Euclid} Data Release 1 \citep{2025arxiv_euclid_i,2025arxiv_euclid_ii,2025arxiv_euclid_iii,2025arxiv_euclid_iv, 2025arxiv_euclid_v}. Previously, \cite{rezaei_machine_2022} trained a CNN on simulated interferometric imaging data from \textit{LOFAR} to classify gravitationally lensed radio sources, using a singular isothermal ellipsoid (SIE) parametrisation with non-zero external shear to generate their simulated training data. \cite{wilde_detecting_2022} showed that CNNs trained on simulated \textit{Euclid} VIS and NISP band imaging are capable of identifying rare compound lens populations -- such as double Einstein rings and compound arcs -- despite not being explicitly trained on them, a result verified on archival \textit{Hubble Space Telescope} (\hst) and Hyper Suprime-Cam observations of known compound systems. Previously, \cite{hezaveh_fast_2017} had used a CNN trained on simulated \textit{HST} imaging, after lens light subtraction, to recover the Einstein radius and ellipticity components of an SIE mass model, demonstrating that parameter inference can be performed approximately ten million times faster than traditional maximum likelihood methods at comparable accuracy, validated on real \hst\ imaging from the Strong Lensing Legacy Survey.

Despite their success, CNNs can be computationally expensive to train and are typically trained on simulated data that, despite best efforts, may not fully represent observed systems, so performance can degrade on real data. \citet{acevedo_barroso_purities_unions_2025} highlight this limitation in the context of computing CNN classifier metrics from simulations rather than real observations. Valuable information is also contained in more compressed representations of the data than the lensed images themselves, such as the observed lensed image positions. \cite{Baltasar2026} recently demonstrated that lens mass model parameters can be inferred from point-source data alone -- namely image positions, fluxes, and time delays -- using the GPU-accelerated Bayesian framework \textsc{GIGA-Lens}. This framework employs a multi-stage pipeline consisting of gradient descent optimisation, variational inference, and Hamiltonian Monte Carlo (HMC) sampling to fully explore the posterior distributions of the lens parameters. While this approach provides a rigorous statistical characterisation of the parameter posteriors, it requires GPU-accelerated computing infrastructure and carries the computational overhead associated with full Bayesian MCMC sampling. In contrast, the method presented here performs parameter inference using a machine learning approach, which provides initial lens model estimates significantly faster and without the need for GPU acceleration, making it well suited for rapid initialisation and large-sample applications.

\subsection{This work}
\begin{figure}
    \centering
    \begin{tikzpicture}[
        node distance=1.4cm,
        box/.style={
            draw,
            rectangle,
            rounded corners,
            align=center,
            minimum width=3.5cm,
            minimum height=0.9cm
        },
        arrow/.style={->, thick, rounded corners},
        failarrow/.style={->, thick, dashed, rounded corners}
    ]
    
    \node[box] (input) {Observed lensed-image \\coordinates and fluxes};
    \node[box, below=of input] (features) {Input feature preprocessing\\(centring, ordering, symmetries)};
    
    \node[box, below left=0.8cm and -1.6cm of features] (coords)
    {Image coordinates};
    
    \node[box, below right=0.8cm and -1.6cm of features] (ratios)
    {Ratios of distances\\between images};
    
    \node[box, below=2.2cm of features] (nn)
    {Neural network regression\\($b^\prime$, $\epsilon$ prediction)};
    \node[box, below=of nn] (est) {Construct initial lens model\\from NN-predicted $b^\prime$, $\epsilon$};
    \node[box, below=of est] (opt) {Lens model optimisation \\(minimise $\chi^2_{\rm pos}$)};
    \node[box, below left=1.2cm and -1.6cm of opt] (fail) {Optimisation failed\\(non-convergent $\chi^2$)};
    \node[box, below right=1.2cm and -1.6cm of opt] (final) {Output mass model};
    
    \draw[arrow] (input) -- (features);
    
    \draw[arrow] (features.south west) -| (coords.north);
    \draw[arrow] (features.south east) -| (ratios.north);
    
    \draw[arrow] (coords.south) |- (nn.west);
    \draw[arrow] (ratios.south) |- (nn.east);
    
    \draw[arrow] (nn) -- (est);
    \draw[arrow] (est) -- (opt);

    \draw[dotted, thick] 
        ([yshift=-0.55cm, xshift=-2cm]est.south west) -- 
        ([yshift=-0.55cm, xshift=2cm]est.south east);
    
    \node[right, font=\footnotesize\itshape, text=gray] 
        at ([yshift=-0.15cm, xshift=0.35cm]est.south east) 
        {Machine Learning};
    \node[right, font=\footnotesize\itshape, text=gray] 
        at ([yshift=-0.95cm, xshift=0.35cm]est.south east) 
        {Optimisation};
    
    \draw[arrow] (opt.east) -| (final.north);
    
    \draw[failarrow] (opt.west) -| (fail.north);
    \draw[failarrow] (fail.east) -| (opt.south);
    
    \end{tikzpicture}
    \caption{Schematic overview of the end-to-end methodology used in this work to infer and optimise gravitational lens mass models.}
    \label{fig:Overall_method_flowchart}
\end{figure}
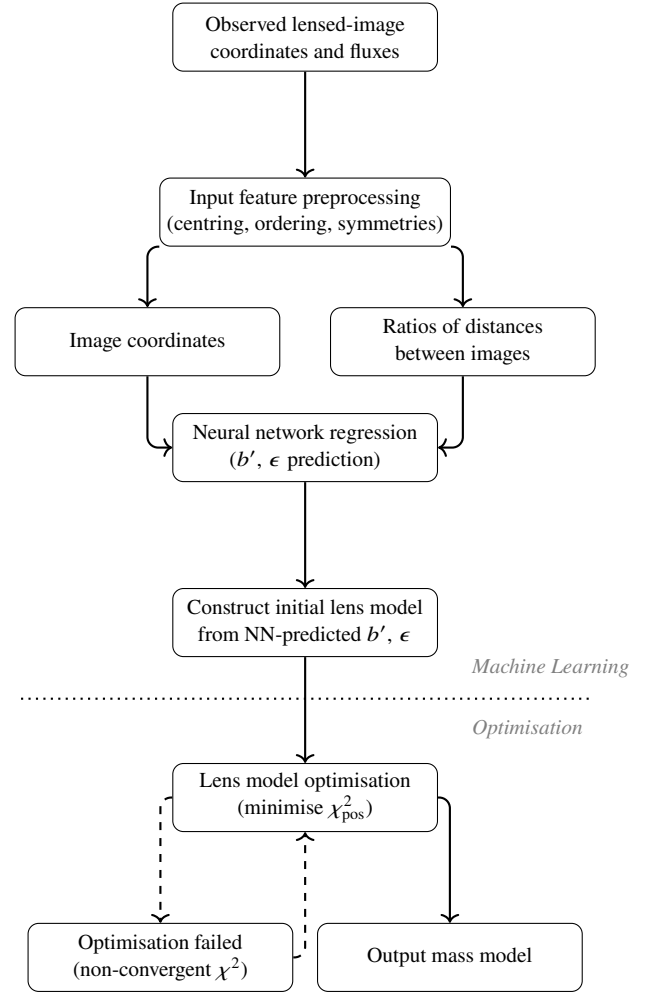

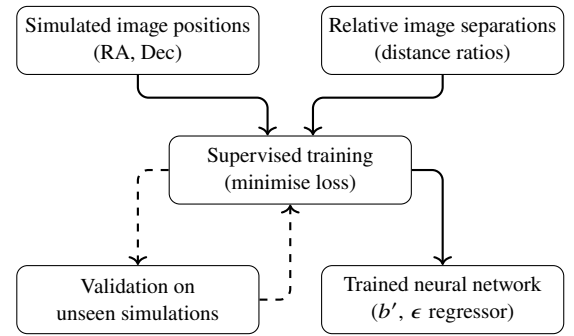
\begin{figure}
    \centering
    \begin{tikzpicture}[
        node distance=1.4cm,
        box/.style={
            draw,
            rectangle,
            rounded corners,
            align=center,
            minimum width=3.2cm,
            minimum height=0.9cm
        },
        arrow/.style={->, thick, rounded corners},
        failarrow/.style={->, thick, dashed, rounded corners}
    ]
    
    \node[box] (train)
    {Supervised training\\(minimise loss)};
    
    \node[box, above left=0.8cm and -1.2cm of train] (img)
    {Simulated image positions\\(RA, Dec)};
    \node[box, above right=0.8cm and -1.2cm of train] (ratio)
    {Relative image separations\\(distance ratios)};
    
    \node[box, below right=0.8cm and -1.2cm of train] (final)
    {Trained neural network\\($b^\prime$, $\epsilon$ regressor)};
    
    \node[box, below left=0.8cm and -1.2cm of train] (test)
    {Validation on\\unseen simulations};
    
    \draw[arrow]
      (img.south) -- ++(0,-0.3) -| ([xshift=-3mm]train.north);
    
    \draw[arrow]
      (ratio.south) -- ++(0,-0.3) -| ([xshift=3mm]train.north);
    
    \draw[arrow] (train.east) -| (final.north);
    \draw[failarrow] (train.west) -| (test.north);
    
    \draw[failarrow]
      (test.east) -| (train.south);
    
    \end{tikzpicture}
    \caption{Schematic of the neural network training workflow.}
    \label{fig:NN_training_method_flowchart}
\end{figure}

We present a machine learning (ML) framework (Fig. \ref{fig:NN_training_method_flowchart}) that accelerates mass modelling for applications of the caustic method in quadruply-imaged point-like systems. Although we illustrate its performance using \gaia\ observations of quadruply lensed AGN, the framework is instrument-agnostic and applies to any facility capable of resolving four point-like images of a lensed source. We propose an alternative ML approach (Fig. \ref{fig:Overall_method_flowchart}) based on relatively simple fully connected neural network architectures that rapidly infer starting values of the mass parameter and ellipticity of lenses parametrised by a singular isothermal ellipsoid (SIE) potential. These estimates can be used as initial values for mass model optimisation for caustic method applications, significantly reducing the computational time required to obtain a high-fidelity model. The SIE model is commonly used to describe the mass distribution of elliptical galaxies, which comprise a significant fraction of known early-type galaxy lenses \citep{lemon_review_2024,schechter_even_2019}. In contrast to ML methods that rely on simulated full image pixel information to perform parameter inference whose training process may be computationally expensive \citep{erickson_2024_neural_post_estimation}, our architectures use minimal information, namely, the observed lensed image positions of quadruply lensed systems.

Our analysis is based on a simulated dataset of $> 6 \times 10^6$ lens-source configurations sampling a broad region of parameter space in mass parameter and ellipticity ($b^\prime = 0\farcs1$--$5\farcs5$, $\epsilon = 0.05$--0.60), which is partitioned into training, testing and validation samples. We show that, in 75 per cent of our SIE-simulated lenses, our novel ML framework provides suitable starting values for the mass and ellipticity parameters so that, when subsequently optimised in time-scales of minutes, the simulated lensed image positions are reproduced to milliarcsecond precision. Across simulated SIE systems spanning our lens configuration parameter space and incorporating randomised ellipticity angles, 151 of 200 cases converged within a reasonable time to mass models capable of reproducing the lensed image positions to 0.5 mas accuracy. Across the same range of simulated SIE systems, we estimate the lens mass parameter with an RMS error of 0\farcs165 and the ellipticity with an RMS error of 0.180. We further validate the framework on seven quadruply lensed quasars observed by \gaia\ and studied in the literature, for which the SIE+shear mass model provides an adequate description.

This paper is organised as follows. In Sec. \ref{sec:Simulations} we discuss the mass model simulations used to train the neural network. In Sec. \ref{sec:ML} we describe the neural network model used to learn the relationship between lensed images and physical parameters, and discuss the hyper-parameters used. In Sec. \ref{sec:results} we describe the results of training the neural network on the simulated data, and present some metrics to evaluate the success of our method. Finally, in Sec. \ref{sec:discussion} we discuss the astrophysical implications of applying our method to real systems.

\section{Simulations of quadruply lensed quasars}

\label{sec:Simulations}
This section summarises the steps and assumptions made to build the simulations used to train the neural network. All simulations were run using the \texttt{lensmodel} routines contained in the \textsc{gravlens} gravitational lensing software package \citep{keeton_software_2004,keeton_modeling_2010}. For a given parametrisation of the effective gravitational potential, \texttt{lensmodel} explores the parameter space of the model's free parameters via chi-squared optimisation of the predicted vs. observed properties of the lensed system. Here, we focus on optimising the model-predicted lensed image positions.

In general, \textsc{gravlens} requires implementing numerical methods to find all the root solutions to the non-linear lens equation \citep{keeton_modeling_2010}. Instead, \texttt{lensmodel} tiles the image plane, and maps these tiles to the source plane. With this, \texttt{lensmodel} calculates the number of lensed images for a given system and estimates their positions; these can subsequently be refined through numerical calculations.

\subsection{Simulations} \label{subsec:simulations}
Of the $\sim 300$ lensed quasars discovered to date \citep{lemon_search_garv_quasars_gaia_i_2023,lemon_search_garv_quasars_gaia_ii_2023}, only $\sim 50$ are quadruply lensed quasars \citep{lemon_review_2024,stern_2021}\footnote{This number is expected to increase substantially with forthcoming data, including \gaia\ Data Release 4 and successive \textit{Euclid} data releases.}. This number is too low to train a neural network. Therefore, we simulated a training set of over 6 million lens-source pairs, simulating each dataset based on the following criteria:
\begin{enumerate}
    \item The mass distribution of all lenses is well-described by a singular isothermal ellipsoid (SIE).
    \item \label{enumerate_assumption} There is no external shear.
    \item All ellipticities ($\epsilon$) are at an angle of 0 radians, where the major axis of the ellipticity is aligned with the declination axis (position angle: $\mathrm{PA} = 0^\circ$).
    \item The lens centre is placed at the coordinate origin, $(\mathrm{RA}, \mathrm{DEC})=(0.0, 0.0)$.
    \item \label{enumerate_assumption_ii} All units and angles are scaled to dimensionless quantities using the formalism in Sec. 1.3 of \cite{keeton_software_2004}.
\end{enumerate}

The SIE mass distribution can be described by a power-law density profile:
\begin{equation}
    \kappa = \frac{1}{2}(b^\prime)^{2-\alpha}(\zeta^2)^{\alpha/2-1}
    \label{eq:powerlaw}
\end{equation}

\noindent where $\zeta$ is the elliptical radius, and $\kappa$ is the dimensionless surface mass density normalised to a critical density. The surface mass density is related to the lens ellipticity via $\zeta$, defined as:
\begin{equation}
    \zeta = [(1-\epsilon)x^2 + (1+\epsilon)y^2]^\frac{1}{2} \text{ with } q^2 = \frac{1-\epsilon}{1+\epsilon},
    \label{eq:epsilon}
\end{equation}
where $q$ is the projected axis ratio. Here, $b^\prime$ is the mass scale parameter of the SIE model as defined in \textsc{gravlens} \citep{keeton_software_2004}, representing the characteristic angular scale of the lens mass in arcseconds. For a spherical model ($\epsilon = 0$), $b^\prime$ reduces to the Einstein radius of the equivalent singular isothermal sphere; for non-zero ellipticity, the true Einstein radius of the SIE depends on both $b^\prime$ and $\epsilon$. In addition, $\alpha$ is the power-law parameter capturing the dependence of the surface mass density of the main lensing galaxy with radius.

In all simulations used in the training set, we assume a value of $\alpha$ = 1, i.e., an isothermal model:
\begin{equation}
    \kappa = \frac{1}{2}b^\prime\zeta^{-1}.
    \label{eq:simplepowerlaw}
\end{equation}

\noindent In addition, as indicated above, we assume no external shear as this parameter is degenerate with the ellipticity of the lensing galaxy and its exclusion reduces the amount of training data required to densely sample the parameter space. The ellipticity angle is set to 0 (aligned with the declination axis) since real data inputs can be rotated to match this constraint. Scaling to dimensionless units and angles renders our approach generally applicable to SIE models for sources and lenses at any redshift and with any lensing mass, since the redshift enters only through the time delays and thus does not affect the image positions on which our method is based. 

\subsubsection{Source Gridding} \label{subsubsec:gridding}
We produce a suite of simulations by scanning through a grid of model parameters (mass parameters and ellipticities). We assume the positions of the source and lensing galaxy to be independent of one another. For each combination of parameters, we compute the resulting lensed image positions for a range of source positions. These computations naturally define two conceptual planes in lens modelling: the source plane, where the background source is located, and the image plane, where the multiple lensed images appear as seen by an observer. 

We simulate mass parameters $b^\prime = 0\farcs1$--$5\farcs5$ in steps of 0\farcs1, and simulate ellipticities $\epsilon = 0.05$--0.60 in steps of 0.05. We compute the resulting lensed image positions for each lens configuration in the grid (i.e. for each $b^\prime$, $\epsilon$ value in the grid). This range of mass parameters is expected to cover the range of values accessible to high resolution X-ray observations \citep{sonnenfeld_sl2s_2015}. Note that \textsc{gravlens} defines the mass parameter as the Einstein radius for a spherical galaxy in arcseconds, with $\epsilon= 0$, following an SIE model. For models with extreme ellipticities, the mass parameter is linearly dependent on the Einstein radius, and can include any positive value. In addition, \cite{chen_ellipticities_2016} identified that the majority of elliptical galaxies have ellipticities $\epsilon \in 0$--0.6. 

The inner caustic of an SIE lensing galaxy with no external shear follows the shape of an astroid with four cusps. Therefore, for each set of $(b^\prime, \epsilon)$ value in the grid, we fit an astroid to the model's inner caustic calculated using \texttt{lensmodel} \citep{keeton_software_2004}, and then place sources along uniformly scaled versions of the fitted astroid in the source plane. Most of the gravitationally lensed sources that we use for X-ray astrometry are located close to the caustics due to the higher magnification resulting in a greater number of identifiable lensed images \citep{barnacka_galaxies_2017,barnacka_gravitational_2018}. 

To sample source positions near the inner caustic, we generate a sequence of astroids centred on the lens that shrink inward. The initial astroid has a semi-major axis
\begin{equation}
a_0 = 0.998 \times a,
\end{equation}
where \( a \) is the semi-major axis of the astroid that touches the caustic. The semi-minor axis is defined using a fixed axis ratio \( R \), such that
\begin{equation}
b = \frac{a}{R},
\end{equation}

\noindent where $b$ is the semi-minor axis of the fitted astroid (not to be confused with the mass parameter $b^\prime$).

\noindent The astroids shrink by reducing the semi-major axis in steps. The initial step size is
\begin{equation}
\delta a_0 = \frac{a_0}{1000},
\end{equation}
and the step size increases at each iteration by a constant factor \( s = 1.2 \):
\begin{equation}
\delta a_{i+1} = s \cdot \delta a_i,
\qquad
a_{i+1} = a_i - \delta a_i.
\end{equation}
This process continues while \( a_i > 0.01 \times a \), producing a sequence of nested astroids approaching the centre.

On each astroid, we place source positions symmetrically. The number of sources \( N \) placed per astroid scales linearly with the semi-major axis:
\begin{equation}
N = \left\lfloor 104 \times \frac{a_i}{a_0} \right\rfloor,
\end{equation}
where 104 is the number of sources placed on the initial astroid.

\begin{figure}
    \centering
    \includegraphics[width=\linewidth]{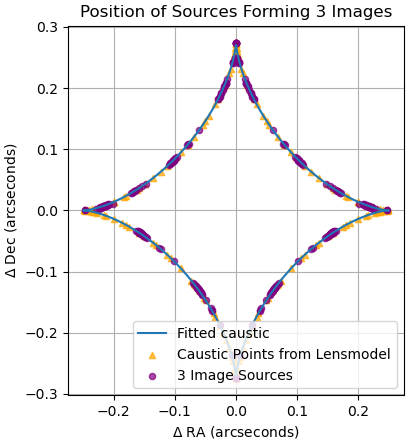}
    \caption{The 278 source positions that produce 3 images are plotted in the source plane in purple for a mass parameter of 2.4, ellipticity of 0.15, with a lens centre (0,0). The inner caustic of the lens, fitted to the output of \texttt{lensmodel}, is plotted in blue with the caustic points calculated in \texttt{lensmodel} in yellow.}
    \label{fig:position3images}
\end{figure}

\begin{figure}
    \centering
    \includegraphics[width=\linewidth]{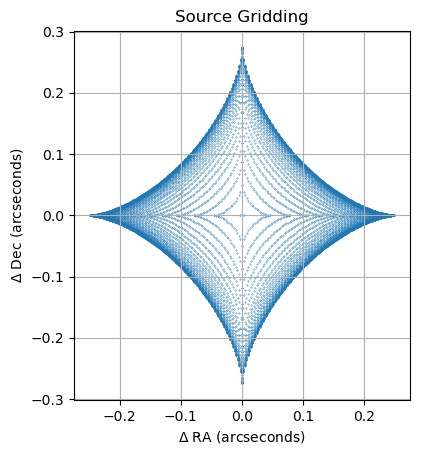}
    \caption{The 10,008 source positions plotted in the source plane for a mass parameter of 2.4, ellipticity of 0.15, with a lens centre (0,0). The number of sources at each distance from the lens centre is decided by a truncated normal distribution.}
    \label{fig:sourcegridding}
\end{figure}

We began placing sources along an astroid scaled to 99.8 per cent of the initial caustic, since at this distance 3 per cent of the sources (278 sources) produce three lensed images rather than four. This is illustrated in Fig. \ref{fig:position3images}. This well-known phenomenon is due to two distinct lensed images being formed so close together that they merge. Sampling further from the caustic produces fewer sources with only three lensed images; however, we want to grid as close to the caustic as possible. We repeat this gridding, getting closer to the lens centre each time, until the whole caustic has been sampled with sources as shown in Fig. \ref{fig:sourcegridding}.
This method of gridding sources produced a grid of over 6 million source--model pairs, which were then evaluated using the software \texttt{lensmodel} to determine their corresponding lensed image positions.

\subsection{Ellipticity Angle and the Lens Centre} \label{subsec:EllipAngleAndLensCentre}
\begin{figure}
    \centering
    \includegraphics[width=\linewidth]{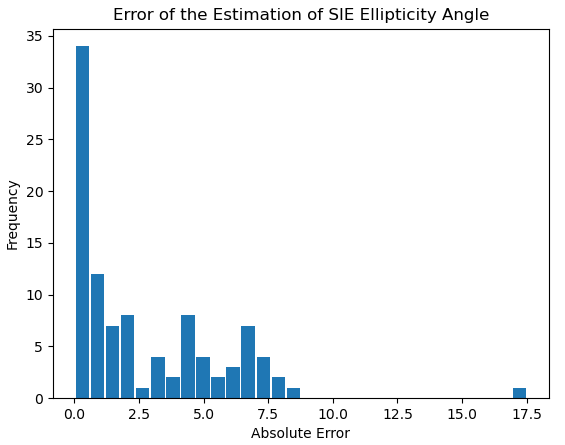}
    \caption{A histogram with 30 bins, displaying the absolute error (in degrees) in the estimated angle of ellipticity of the simulated lenses, for a batch of 100 random lens configurations and source positions.}
    \label{fig:EllipAngleError}
\end{figure}

In general, SIE lenses are parametrised by their mass, ellipticity, ellipticity angle, and lens centre. Their properties are also strongly influenced by their fixed power-law density slope ($\alpha=1$ in Eq. \ref{eq:powerlaw}). In general, lenses will be observed with different orientations; however, we simulate only an ellipticity angle of 0 (where the major axis of the ellipse is aligned with the declination axis). This choice is made for simplicity, as it has no discernible effect on the underlying physics. Training a fully connected neural network on random ellipticity angles would substantially increase both the size of the training dataset and the computational time, while also complicating the convergence of the model. To address this, we estimate the orientation of the ellipticity using the positions of the lensed images and the centre of the lens, assuming a quad-lensed source. 
\cite{kassiola_invariants_1995} showed that the position angle $\theta$ of a singular isothermal lensing galaxy producing a quad can be calculated from the position angles $\theta_\mathrm{i}$ $(i=1, 2, 3, 4)$ of the four lensed images relative to the lens centre:

\begin{equation}
    \theta = \frac{1}{4}(\theta_{1} + \theta_{2} + \theta_{3} + \theta_{4}) \pm \frac{\pi}{4} \label{EllipApprox}
\end{equation}
assuming that the lens centre is known.
With no external shear in our simulations (Sec. \ref{subsec:simulations}), the orientation of the lens is set entirely by the galaxy's ellipticity, so this recovered position angle is the ellipticity angle of the SIE. For real systems which may contain external shear, we use this angle as an initial estimate since ellipticity and shear are degenerate.
Each angle $\theta_{i}$ is taken anti-clockwise between the positive RA-axis and the image position with the mass centre of the lens at the origin. Given that the mass centre is unknown for most real lenses, we calculate $\theta_{i}$ from the intersection point of opposing image positions. The $\pm \pi/4$ signifies that it is not explicitly known which angle is the major/minor axes for a specific set of images and so both values are used in the final optimisation of the mass model. 
After testing on 100 random simulated SIE lens configurations and image positions with random rotations, we calculate the error between the best ellipticity rotation estimation (either major or minor since both axes are used during optimisation) and the true values. A histogram of errors for such randomly simulated lens configurations is shown in Fig. \ref{fig:EllipAngleError}. The mean error for the estimation of the major/minor axes was 2.73 degrees. This ellipticity angle estimate is within an acceptable range for the majority of cases such that varying the ellipticity angle during optimisation should lead to an accurate final mass model.

\subsubsection{Lens Centre} \label{subsec:lenscentre}
To fully describe the lens, we require a good estimate of the position of the lens centre. We can approximate the lens centre position as the centre of an ellipse fitted to the four image positions due to the central symmetry of the lensing potential in SIE models, which causes the images to approximately trace an elliptical critical curve around the lens centre. To evaluate the accuracy of this approximation, we minimised the sum of the squared deviations between image positions and an ellipse using the Python \texttt{scipy.optimize.minimize} module. We then compared the centre of the best-fitting ellipse to the true lens centre for the mass model, using 100,000 randomly generated lens configurations and source positions drawn from our test dataset. The root mean squared error (RMS) of the estimated ellipse centre was 1\farcs284 in Right Ascension and 1\farcs279 in Declination.

Alternatively, we could have found the mass centre of the lensing galaxy using the analytic calculation of \cite{schechter_even_2019}. Using an SIE potential and assuming that there is no $\epsilon$ component perpendicular to the ellipticity angle, the authors demonstrated that, for a quad system, the four resulting lensed images lie along an ellipse centred on the source position. The resulting image configuration is fitted by a hyperbola whose asymptotes are aligned with the axes of the ellipse. This method reduces the dimensionality of the parameter space by constraining the positions of the lens and source and the orientation of the system, simplifying the model fitting process. However, this exact analytical solution is limited to SIE lenses where the external shear is aligned with the principal angle of the main deflector. 
In practice, the mass centre of the lens may be measured directly from observations if the lensing galaxy is resolved. This could provide more accurate results compared to using the ellipse estimation. However, our machine learning approach is intended to be applied to observed lensed image positions from \gaia, where the lensing galaxy is typically not resolved, and thus falls beyond the scope of this paper.

\section{Machine Learning Methodology}
\label{sec:ML}
This section presents the key aspects of the machine learning architecture we have employed, as well as its main inputs and outputs. First, we describe the design of our neural network (Sec. \ref{subsec:networkarchitecture}) and the input features, including the feature selection process and the feature categorisation method (Sec. \ref{subsec:networkfeatures}). Second, we outline the training process, focusing on the convergence criteria and how the networks are trained (Sec. \ref{subsec:networktraining}). Third, we discuss the mass model optimisation procedure, detailing the method used to refine the high-fidelity models to reproduce lensed image positions (Sec. \ref{subsec:optimisation}).

\subsection{Network Architecture} 
\label{subsec:networkarchitecture}
We deliberately adopt a simple neural network architecture, but train two separate models: one to predict the mass parameter, and the other to predict the ellipticity. Both models use the same set of input features. While a single multi-output network to predict both parameters simultaneously was initially explored, we found that treating the mass parameter and ellipticity as independent regression tasks consistently improved prediction accuracy for both quantities. A quantitative comparison of the two approaches is provided in Table \ref{tab:NNComparison}, where we report RMS errors for both the joint and separate network configurations. The separate networks achieve an RMS error of 0\farcs1481 and 0.0299 for the mass parameter and ellipticity respectively, compared to 0\farcs1778 and 0.0941 for the joint network, demonstrating a consistent improvement in accuracy across both parameters, which motivated our choice of independent regression models for the final architecture. Each model is a fully connected, feed-forward neural network with two hidden layers. The first hidden layer contains 100 neurons, while the second contains 40 and 50 neurons for the mass-parameter and ellipticity networks, respectively. All hidden neurons use a rectified linear unit (ReLU) activation function with a dropout percentage of 20 per cent (see discussions in Appendices \ref{appendixLossFunction} and \ref{appendixReLU}). The input layer consists of a set of features derived from the coordinates of the four lensed images, and the output layer has a single neuron with a linear activation function, to predict either the mass parameter or the ellipticity of the lens.

\begin{table}
    \centering
    \caption{Comparison of prediction accuracy for a joint multi-output network and separate single-output networks for the mass parameter $b^\prime$ and ellipticity $\epsilon$. The root mean squared error (RMS) is reported for each configuration, evaluated on 1,000 randomly selected lens configurations from the test dataset. The separate networks achieve lower RMS errors for both parameters compared to the joint network, motivating the use of independent regression models in our final architecture.}
    \begin{tabular}{c|c|c|c}
        Network Outputs & Network Hidden Layers & $b^\prime$ RMS & $\epsilon$ RMS \\
        \hline
        $b^\prime$, $\epsilon$ & 100 -- 50 & 0\farcs1778 & 0.0941\\
        $b^\prime$ & 100 -- 40 & 0\farcs1481 & -- \\
        $\epsilon$ & 100 -- 50 & -- & 0.0299 \\
    \end{tabular}
    \label{tab:NNComparison}
\end{table}

This architecture results in a system with a total of 5,581 and 6,601 trainable parameters for the mass parameter and ellipticity networks respectively. Neural networks such as the ones we have described are universal approximators. The ReLU activations create piece-wise segments of a multivariate function that approximate the relation between the input features and the target quantities on which we regress. By training this system, we find the values of the network parameters that minimise the error between the predicted relationship and the training data. 

\subsection{Input Features} 
\label{subsec:networkfeatures}
In order to predict the lens parameters accurately, the neural network requires informative inputs that can be inferred from observations and that describe the geometry of the lensing system. We begin by considering several observables that are available from our simulated observations: 

\begin{enumerate}
\item the positions of each of the lensed images (where each lensed image is defined by its corresponding RA, DEC) relative to the intersection point of pairs of opposite images\footnote{Opposite images refers to non-adjacent images, the lensing galaxy typically lies between each such pair \citep{luhtaru_what_2021}.}, \label{input_features:image_pos}
\item the distance ratios between lensed images in the image plane, \label{input_features:distance_ratios}
\item the distances between the lensed images in the image plane in arcseconds, 
\item the flux ratios between lensed images, \label{input_features:flux_ratios}
\item the areas enclosed by each set of three lensed images. \label{input_features:area_tri_enclosed}
\end{enumerate}

\noindent These parameters fully describe the observed image positions and fluxes in the absence of flux variability. However the observed fluxes are expected to vary, both due to intrinsic quasar variability and due to microlensing by stars in the lensing galaxy \citep[comprehensively reviewed in][]{vernardos_microlensing_rev_2024}. Feature \ref{input_features:flux_ratios} may therefore be invalidated (Sec. \ref{subsubsec:resultsMicrolensing}). All input features are constructed solely from the four image positions and their associated fluxes. However, the direct use of absolute image positions requires a constant and physically motivated reference frame across different lens systems. To ensure this, we define the origin of the coordinate system as the intersection point of pairs of opposing images, providing a translation-invariant reference frame that can be applied uniformly to all systems.

To assess the importance of the available features, we trained the neural network on every possible pair of features among the five candidate features listed above, and subsequently evaluated their performance in predicting the mass and ellipticity parameters. Network performance was quantified using the mean squared error (MSE) of the predicted parameters. From this comparison, we found that the combination of lensed image position \ref{input_features:image_pos} and ratios of the separations between image pairs \ref{input_features:distance_ratios} consistently yielded the lowest prediction errors. We therefore adopt these two feature sets as inputs for all subsequent training and analysis, and use them exclusively to predict both the mass parameter and ellipticity.

As neural networks train on pure numerical input values and apply numerical operations to the inputs in a fixed order, the input features should be provided in the same order for all examples to keep the relationships between nodes consistent. Otherwise, with randomised inputs between all input nodes, there will be no patterns or relationships for the network to learn. 
This work used the following labelling method for the lensed images:
\begin{itemize}
    \item \textbf{Image A} labelled as the lensed image with the highest model-predicted absolute flux.
    \item \textbf{Image B} labelled as the lensed image closest to image A.
    \item Consecutively label \textbf{images C \& D} in either a clockwise or anti-clockwise direction, starting from B, based on the direction of B relative to A. 
\end{itemize}
All model-predicted image fluxes are dependent on the model-predicted magnifications, and are thus sensitive to the location of each lensed image relative to the source position. The fluxes of each lensed image can be altered by microlensing and by the presence of additional mass deflectors along the line of sight that perturb the gravitational potential; however, since flux is used only to identify image A and is not itself a network input, such effects can affect the predictions only indirectly, by altering the image labelling for a given system at a given epoch. Our simulations include no such effects, so the labelling and therefore the parameter predictions are exact here; we discuss the real-data case in Sec. \ref{subsubsec:DiscussMicro}.

\subsection{Network Training} \label{subsec:networktraining}
We shuffled and split the dataframe of over 6 million source-model pairs (where each `model' denotes a specific lens configuration), with 70, 20 and 10 per cent attributed to the training, testing and validation sets respectively. We trained all the models using the \texttt{tensorflow} library in \textsc{Python} using the training dataset. The mean squared error of the predicted parameters was used as the loss function as it heavily penalises outlier predictions. The validation loss (the loss evaluated on the validation set) was recorded after each epoch. We used the Adam optimiser with its default learning rate of 0.001 to compute the back-propagation updates. As mentioned earlier, each layer of the network was followed by a dropout layer with a dropout rate of 20 per cent to improve generalisation.

All models that we train have an early stopping criterion with a maximum of 50 epochs with a validation loss improvement of less than $10^{-4}$ per epoch. This approach not only helps prevent overfitting but also enables efficient testing of a wide range of neural network architectures, e.g. different input features, output parameters and layer configurations. We originally trained a single neural network to predict the mass and ellipticity parameters simultaneously; however, the most accurate parameter predictions were achieved when each parameter was predicted by a separate network architecture. 

\subsection{Model optimisation} \label{subsec:optimisation}
After the NNs predicted the mass and ellipticity parameters, we refined these initial estimates using an optimisation routine within \texttt{lensmodel}. The goal was to obtain a model that reproduces the observed image positions in the image plane to within the astrometric precision required by the caustic method described in Section \ref{Sec:Introduction}. In each optimisation run, the software varied a subset of mass-model parameters. The SIE+shear mass model contains 10 parameters \citep[][]{keeton_catalog_2002}; for the softened power-law model used here, 3 parameters are not optimised. These three parameters are: a core radius fixed to $s' = 0$, the power-law index to $\alpha = 1$, and a second scale radius that the model does not use. Setting $\alpha = 1$ sets a singular isothermal ellipsoid mass distribution -- where the total mass density is $\rho \propto r^{-2}$. The remaining 7 parameters of each optimisation run -- specifically the lens centre position, the mass and ellipticity parameters of the lens, the external shear, and their associated position angles -- constitute the full set of parameters available for optimisation. The procedure began conservatively, with only the lens centre position initially allowed to vary. In subsequent runs, the set of free parameters was progressively expanded to explore parameter space more fully and to improve the model-predicted lensed image positions.

For each optimisation run beyond the initial iteration, up to three parameters are randomly selected from the seven available and varied simultaneously. This limit balances model flexibility against computational stability, as permitting a larger number of free parameters can result in slow or unstable convergence. If a run exceeded 20 seconds without converging, measured on the reference hardware described in Appendix \ref{appendix:computational_setup}, the selected parameter subset was discarded and a new random subset was chosen before restarting the optimisation.

The iterative optimisation process is run until the model reproduces a target astrometric accuracy, defined by a $\chi^{2}$ threshold \eqref{eq:chinumbers}. This $\chi^2$ statistic is defined as: 

\begin{eqnarray}
    \chi^{2} &=& \sum_{i=1}^{4}
    \frac{(\text{Pred. Image Pos.}_i - \text{Measured Image Pos.}_i)^{2}}
    {(\text{Measured Image Pos. Error})^2}
    \label{eq:chi} \\
     &=& \sum_{i=1}^{4}
     \frac{(x_{pred,i} - x_{meas,i})^{2} + (y_{pred,i} - y_{meas,i})^{2}}
     {(\sigma_{meas})^2}, 
     \label{eq:chinumbers}
\end{eqnarray}

\noindent determining the level of agreement between model-predicted and observed image positions. The threshold $\chi^{2} < 8$ is adopted as a pragmatic stopping criterion, corresponding to a target astrometric accuracy of 0.5 milliarcseconds per image coordinate across all four lensed images. It should be noted that this threshold does not correspond to the statistically optimal reduced chi-squared ($\chi^{2}_{\rm red} = 1$), where $\chi^{2}_{\rm red} = \chi^{2} / \nu$, and $\nu = N_{\rm data} - N_{\rm free}$ is the number of degrees of freedom, with $N_{\rm data}$ the number of observational constraints and $N_{\rm free}$ the number of free parameters. Achieving $\chi^{2}_{\rm red} = 1$ would therefore require $\chi^{2} = \nu$. Rather, $\chi^{2} < 8$ serves as a pragmatic upper bound corresponding to a target astrometric accuracy of 0.5 milliarcseconds per image coordinate across all four lensed images: models satisfying this criterion have not necessarily reached the global minimum, and can in principle be further refined to achieve $\chi^{2}_{\rm red} \sim 1$ (i.e. $\chi^{2} = \nu$) through extended optimisation. If the model fails to improve over 10 consecutive attempts, the optimisation is restarted using the alternative candidate ellipticity angle \eqref{EllipApprox} (Sec.~\ref{subsec:EllipAngleAndLensCentre}), since the image positions alone cannot determine which of the two candidates is the major axis.

\noindent We acknowledge that because the number of free parameters varied at each optimisation step (up to 3 from a set of 7) is not fixed across systems, the effective degrees of freedom are not uniform, and a single threshold $\chi^{2} < 8$ is therefore not strictly equivalent to $\chi^{2}_{\rm red} < 1$ in all cases. Nonetheless, this threshold provides a consistent and computationally practical convergence criterion across the full sample. A more rigorous criterion would be to define an adaptive $\chi^{2}$ target at each optimisation step based on the number of free 
parameters active at that iteration; this is noted as a direction for future improvement. Furthermore, we note that the adopted threshold is specific to \gaia's astrometric precision of 0.5 milliarcseconds; if a different survey instrument were adopted, the convergence criterion would need to be revised accordingly.

\begin{figure}
    \centering
    \includegraphics[scale=0.35]{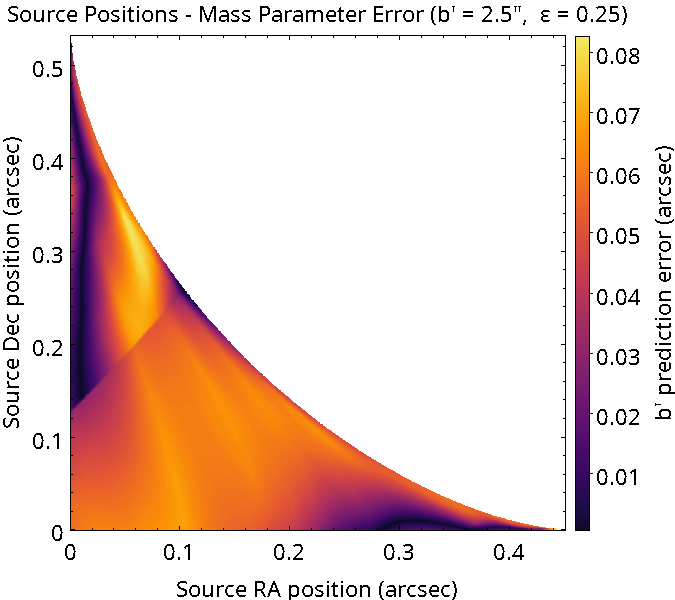}
    \caption{The absolute error on the mass parameter predicted by the neural network for 1,000,000 unseen source positions generated from a uniform distribution across the inner caustic of an SIE lens configuration with a mass parameter of 2\farcs5 and ellipticity 0.25. Sources are placed up to within 0\farcs01 of the caustic to avoid producing 3 images due to \texttt{lensmodel} software limitations.}
    \label{fig:sourceprederrorM25}
\end{figure}

\begin{figure}
    \centering
    \includegraphics[scale=0.35]{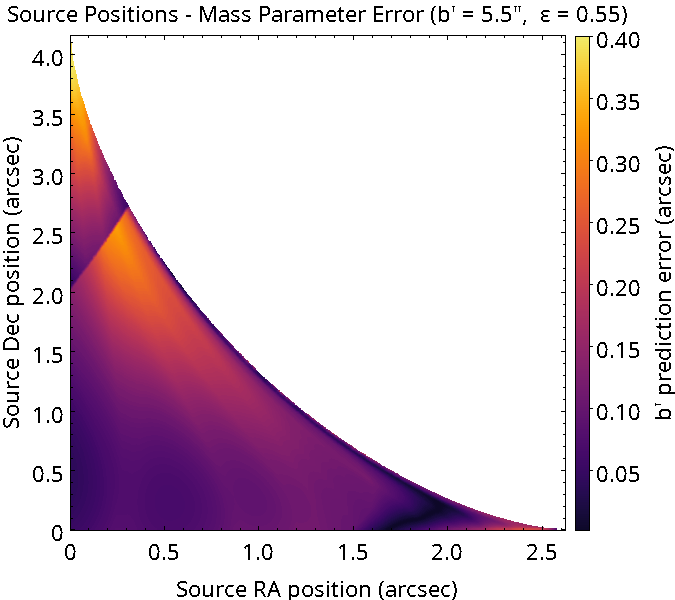}
    \caption{The absolute error on the mass parameter predicted by the neural network for 1,000,000 unseen source positions generated from a uniform distribution across the inner caustic of an SIE lens configuration with a mass parameter of 5\farcs5 and ellipticity 0.55. Sources are placed up to within 0\farcs01 of the caustic to avoid producing 3 images due to \texttt{lensmodel} software limitations.}
    \label{fig:sourceprederrorM55}
\end{figure}

\begin{figure*}
    \centering
    \includegraphics[scale=0.57]{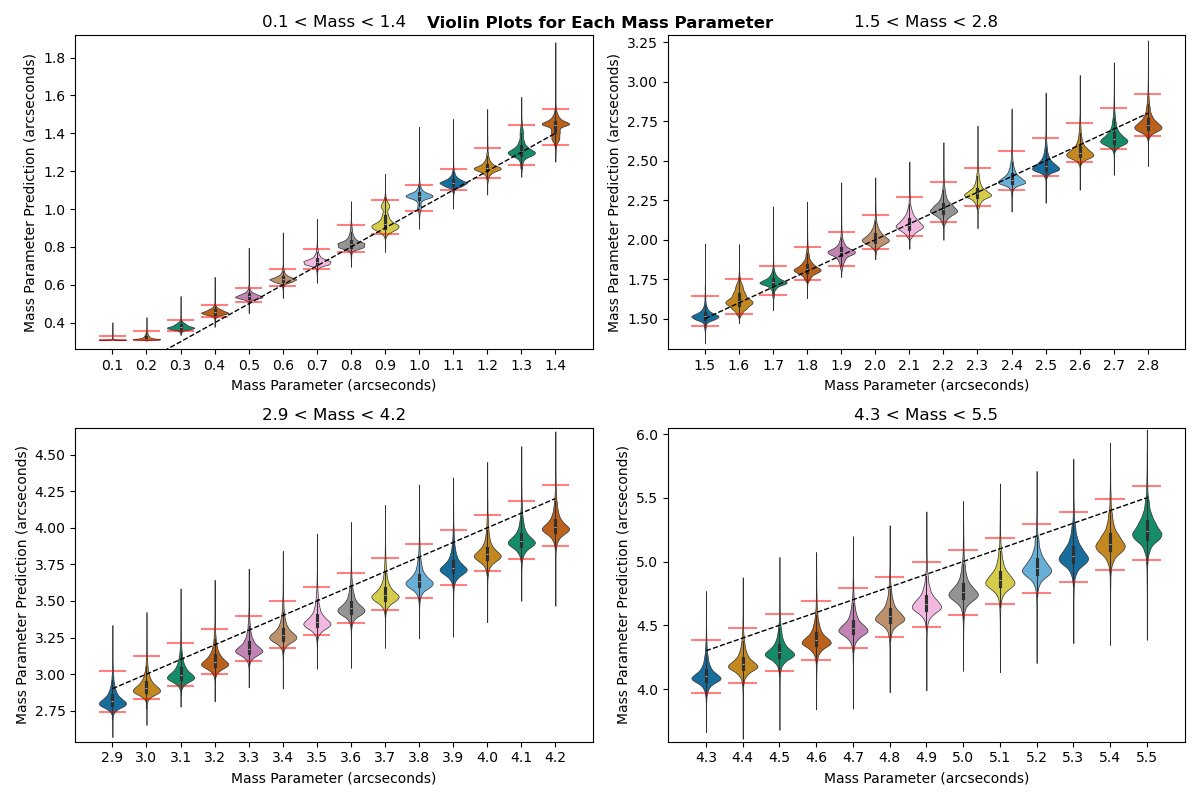}
    \caption{Violin plots of the predicted mass parameter distribution for all testing data sources (1.2 million) across the 55 simulated mass parameters. A dashed identity line is plotted highlighting the true mass parameter values. The width of each violin is fixed and therefore cannot be compared. The red lines on the plots mark the $2.5^{\text{th}}$ and $97.5^{\text{th}}$ percentiles for each mass parameter.}
    \label{fig:massViolin}
\end{figure*}

\begin{figure*}
    \centering
    \includegraphics[scale=0.57]{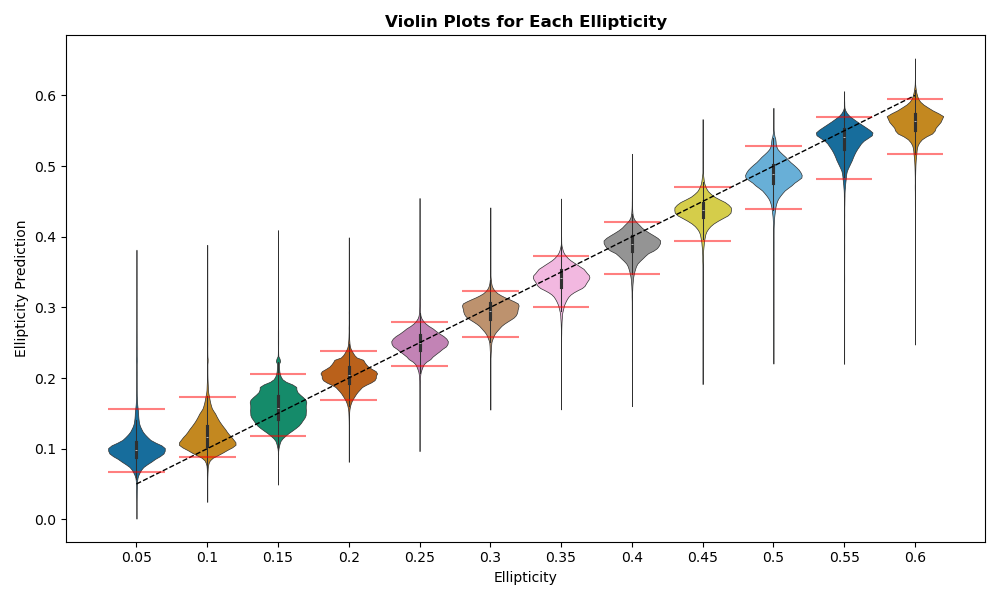}
    \caption{Violin plots of the predicted ellipticity distribution for all testing data sources (1.2 million) across the 12 simulated ellipticities. A dashed identity line is plotted highlighting the true ellipticity values. The width of each violin is fixed and therefore cannot be compared. The red lines on the plot mark the $2.5^{\text{th}}$ and $97.5^{\text{th}}$ percentiles for each ellipticity.}
    \label{fig:ellipViolin}
\end{figure*}

\section{Results} \label{sec:results}
\subsection{Lens model parameter predictions}
The mass and ellipticity parameters were trained on separate neural networks with different architectures (Sec. \ref{subsec:networktraining}). To evaluate the network performance after training, we computed the mean error between the predicted values and their ground truth values for 1,000 randomly selected simulated source-lens configuration pairs that the network had never seen. These random lensed systems had no ellipticity rotation, so the ellipticity was already aligned with the training set. For real data, the ellipticity angle would have to be calculated, and the image positions rotated to match the training set alignment. The results are summarised in Table \ref{tab:NeuralNetworkResults}.

\begin{table}
    \centering
    \caption{Neural Network training results, tested on 1,000 simulated lens configurations from the testing dataset.}
    \begin{tabular}{c|c|c|c}
        Network Trained & Training Epochs & Percentage Error & RMS \\
        \hline
        Mass Parameter & 171 & 9.72\% & 0\farcs1481 \\
        Ellipticity & 641 & 16.52\% & 0.0299 \\
    \end{tabular}
    \label{tab:NeuralNetworkResults}
\end{table}

Figures \ref{fig:sourceprederrorM25} and \ref{fig:sourceprederrorM55} detail the error in predicting the mass parameter for 1 million random source positions in a quarter of the inner caustic for two lens configurations with mass parameters 2\farcs5 and 5\farcs5, and ellipticities 0.25 and 0.55 respectively. Both models have their ellipticity angle aligned with the declination axis, and do not include external shear. The error in ellipticity and mass parameter predictions varies significantly for different lens configurations of the lensing galaxy, and also as a function of the true lensed source position. This variation is likely due to the denser sampling of source positions near the caustics in the training data, where image position and magnification undergo significant changes, leading to more accurate predictions in these regions and greater uncertainty elsewhere. In addition, a sharp transition in prediction accuracy is observed across all quadrants of the caustic for both mass and ellipticity predictions. This transition manifests as a discontinuity in the prediction error, arising from a labelling ambiguity in the image ordering scheme: as the source crosses this labelling discontinuity line within the caustic interior, the relative brightness ordering of the two most highly magnified images changes, producing two geometrically distinct input states for the same lens configuration. Since the network cannot distinguish between these two cases from the input coordinates alone, the conflicting mappings learned during training produce a sharp transition in prediction accuracy, which we discuss further in Sec.~\ref{subsubsec:DiscussMass}.

\subsection{Optimisation}
A mass model is considered optimised when it can predict the image positions within the accuracy to which the input image positions have been measured. We use a value of 0.5 milliarcseconds as the expected measured accuracy, based on the astrometric precision of \gaia. Due to relatively few lensed quasars with observed image positions, we test the results on simulated data that the trained neural networks have never seen -- 200 lens configurations with randomly generated ellipticity rotation angles -- and follow the method laid out in Sec. \ref{subsec:optimisation}. These 200 lens configurations are separate from the testing dataset as we also simulate random ellipticity angles. All optimisation procedures and associated runtimes we report were obtained using the computational setup described in Appendix \ref{appendix:computational_setup}.

Of the 200 lens configurations, 151 reached a chi-squared of $< 8$ and were able to reproduce the image positions to an average error of 0.5 milliarcseconds. They took a mean time of 126.1 seconds for each complete optimisation process, with a mean time of 166.7 seconds for optimisation processes that failed. Convergence was halted and the optimisation restarted when a model -- using the initial mass parameter and ellipticity estimates -- failed to reproduce the known image positions within the specified accuracy (0.5 milliarcseconds). Models that failed to improve after multiple restart attempts were subsequently terminated. When optimising for each model, every optimisation attempt was allowed to run for 20 seconds before being timed out and re-attempted with a different set of variable lens parameters. If a model was timed out 10 consecutive times, the optimisation process was terminated. This was done to avoid getting stuck in a local minimum of the parameter space during convergence and to terminate failed models early, avoiding unnecessarily long convergence times. 

Of the 49 lens configurations that did not achieve $\chi^{2} < 8$, we identify three distinct failure populations. Six systems returned
$\chi^{2} \sim 10^{11}$, the maximum value reported by \texttt{lensmodel}, indicating complete optimisation failure. The remaining 43 systems returned finite but large $\chi^{2}$ values, with a median of $\sim 4 \times 10^{5}$, confirming that these are not borderline cases marginally above the convergence threshold but rather genuine optimisation failures. Three systems failed to improve after exhausting all restart attempts, exceeding 
800 seconds runtime each, and were subsequently terminated; of these, 1 returned $\chi^{2} \sim 10^{11}$ and is included in the complete failure population, while the remaining 2 returned finite $\chi^{2}$ values and are included in the second population. The 151 successfully converged systems achieved a mean $\chi^{2} = 4.957$, a median of $5.391$, and a range of $0.070$--7.995, confirming that converged systems are well within the target threshold of $\chi^{2} < 8$. Table \ref{tab:optimisationResults} lists the resulting data from optimising the 200 models.

\begin{table}
    \centering
    \caption{Results from the optimisation of 200 randomly simulated lensed sources, with randomly generated ellipticity rotation angles. The RMS for all mass parameter and ellipticity optimisation attempts was 0\farcs165 and 0.180 respectively.}
    \begin{tabular}{c|c|c}
        Optimisation Results & $\chi^{2} < 8$ & $\chi^{2} > 8$ \\
        \hline
        Number of Lensed Systems & 151 & 49 \\
        Smallest Image Separation $< 0\farcs5$ & 28 (77.78\%) & 8 (22.22\%) \\
        Smallest Image Separation $> 0\farcs5$ & 123 (75.00\%) & 41 (25.00\%) \\
        Error on Mass Centre $< 0\farcs5$ & 75 (78.95\%) & 20 (21.05\%) \\
        Error on Mass Centre $> 0\farcs5$ & 76 (72.38\%) & 29 (27.62\%) \\
        Time to Complete $<$ 180 sec & 133 (76.88\%) & 40 (23.12\%) \\
        Time to Complete $>$ 180 sec & 18 (66.67\%) & 9 (33.33\%) \\
        Mass Prediction RMS Error & 0.149 & 0.206 \\
        Ellipticity Prediction RMS Error & 0.145 & 0.260 \\
    \end{tabular}
    \label{tab:optimisationResults}
\end{table}

To characterise whether non-convergence clusters in specific configurations, we examined the optimisation outcome ($\chi^{2} < 8$ vs. $\chi^{2} > 8$) using three diagnostics available for all 200 systems: the smallest projected image separation, the positional offset between the recovered and true mass centre, and the wall-clock time required to reach the solution.
We find that the failure rate is insensitive to image separation; systems with the smallest separations ($<0\farcs5$) fail at 22.22 per cent, similar to the 25.00 per cent failure rate for more widely separated systems. Therefore, non-convergence is not simply a consequence of tightly packed image configurations, and does not appear confined to one region of image-plane geometry.
Two other diagnostics show a modest trend: systems with a larger mass-centre offset ($>0\farcs5$) fail at 27.62 per cent versus 21.05 per cent for well-centred systems, and systems requiring more than 180 seconds to complete fail at 33.33 per cent versus 23.12 per cent for faster systems. Both indicate that the harder a system is for the optimiser to localise and refine, the more likely it is to exhaust the iteration budget before reaching $\chi^{2} < 8$.

To evaluate the applicability of our approach to real data, we applied it to a sample of seven quadruply lensed quasars whose four lensed images are observed by \gaia. These systems span a range of environments, from near-isolated lensing galaxies to galaxies with a nearby perturber or group-scale external shear; in all cases we model the main deflector as a single SIE and optimise for a shear component, which suffices to reproduce the image positions to the required accuracy. The optimised mass models and the time taken to converge for each quad are shown in Table \ref{tab:RealImages}. The optimisation was halted once the $\chi^{2}$ stopping criterion of $< 8$ was satisfied, corresponding to reproduction of the observed lensed image positions to within 0.5 mas \eqref{eq:chinumbers}. These models are therefore not necessarily fully optimised and could be refined further through extended optimisation. We also note that the optimisation considers image positions alone without incorporating time delays or flux ratios. The resulting mass models should not be regarded as complete for analyses that demand the high-precision characterisation of the lens potential required for time-delay cosmography, particularly in contexts where the mass-sheet degeneracy must be rigorously constrained \citep{Falco_1985, Schneider_2013}.

\begin{table}
    \centering
    \caption{Optimisation results for an SIE+shear mass model inferred with our ML framework using the observed positions of the four lensed images by \gaia\ for seven individual quadruply lensed quasars in the literature. The model parameters of the SIE+shear model are: the mass parameter ($b^\prime$), the ellipticity ($\epsilon$), and external shear ($\gamma$). For each quad, the optimisation stops when the global chi-squared value of the model-predicted (RA, DEC) for each lensed image against their observed values falls below 8.}
    \begin{tabular}{c|c|c|c|c|c}
        Source & $\chi^{2}$ & $b^\prime$ & $\epsilon$ & $\gamma$ & Time (s) \\
        \hline
        HE0435-1223 & 2.504 & 1.200 & 0.0211 & 0.0721 & 22.34 \\
        J014710+463040 & 7.880 & 1.862 & 0.1835 & 0.1906 & 68.46 \\
        B1422+231 & 7.989 & 0.755 & 0.4359 & 0.1552 & 210.60 \\
        J0659+1629 & 6.468 & 2.454 & 0.0820 & 0.1002 & 108.61 \\
        2M1134-2103 & 3.071 & 1.183 & 0.6167 & 0.1933 & 45.53 \\
        PG1115+080 & 7.866 & 1.042 & 0.5668 & 0.0798 & 318.91 \\
        WFI2033-4723 & 7.308 & 0.884 & 0.8218 & 0.5008 & 218.16 \\
        
    \end{tabular}
    \label{tab:RealImages}
\end{table}

\subsubsection{Microlensing} \label{subsubsec:resultsMicrolensing}
Microlensing is not used as a direct network input, as flux ratios are excluded. However, it can affect the initial image labelling (Sec. \ref{subsec:networkfeatures}) by altering which image is brightest, which in turn influences the $b^\prime$ and $\epsilon$ predictions. This has no effect on our simulated systems, where fluxes are exact, but could in principle affect the real systems considered here; we discuss this further in Sec. \ref{subsubsec:DiscussMicro}.

\section{Discussion} \label{sec:discussion}
Here, we first discuss the performance of the neural networks on real gravitationally lensed sources (Sec. \ref{subsec:reallensedsources}), and on the simulated SIE models (Sec. \ref{subsec:MassModelOpt}). We then discuss the errors on the predictions of mass parameter and ellipticity, and the effect of microlensing on these predictions (Sec. \ref{subsubsec:ParameterPredictions}). Finally, we discuss how error propagation is currently handled and possible avenues for improvement (Sec. \ref{subsec:DiscussPropErrors}).

\subsection{Observed Lensed Sources} \label{subsec:reallensedsources}
Table \ref{tab:RealImages} shows mass model optimisations of seven quadruply lensed quasars whose four lensed images were observed by the \gaia\ Space Observatory. Given that current X-ray observations of image positions have a resolution of 0\farcs5, we attempt to reproduce the images to the accuracy of optical data from \gaia, which has an astrometric uncertainty of $<0\farcs001$ for lensed quasar images \citep{Lindegren_2021_astrometric, Lindegren_2021_parallax}. All seven systems converged to 
$\chi^{2} < 8$ for their predicted image positions and reproduced the observed image positions to within $\pm$0.5 milliarcseconds. The optimisation process took between 22 and 319 seconds per source, with a mean convergence time of 142 seconds. However, all models were halted in their optimisation process after reaching a $\chi^{2}$ value lower than 8. The process could be continued if a greater level of precision were required.

We compare our recovered lens configurations against published models for several systems in our sample. For HE0435-1223\footnote{The H0LiCOW/TDCOSMO programme measured the Hubble constant from time-delay cosmography using a sample of six lensed quasars \citep{Wong_2020}, three of which are included in our sample: HE0435-1223, PG1115+080, and WFI2033-4723.}, our recovered mass parameter $b^\prime = 1.20$ agrees closely with the Einstein radius of $1\farcs20$ obtained from a smooth SIE+shear fit by \citet{nierenberg_2017}, and lies within the range of published values inferred under different treatments of the nearby companion galaxy. We note that $b^\prime$ is the \texttt{lensmodel} mass normalisation and is not in general equal to the Einstein radius, differing by an ellipticity-dependent factor \citep{keeton_catalog_2002}. For this near-circular system ($\epsilon$ = 0.02), the two parameters differ by less than one per cent, so the comparison holds. Published models for PG1115+080 account for its local environment \citep{Wong_2020}. Our single SIE+shear configuration does not capture such structure, but it is not required to: for the purpose of X-ray astrometry it need only reproduce the observed image positions, which it does to within the required tolerance (Table \ref{tab:RealImages}). The same holds for J0659+1629: in \citet[][who modelled the system based on the positions of its four \gaia\ DR3 lensed images]{siskreynes_2026}, the authors modelled the main lensing galaxy together with a perturber next to one of the lensed images (with initial positions taken from \hst\ data); our simpler configuration nonetheless reproduces the observed positions to the required astrometric accuracy.

It is important to note that multiple lens configurations can reproduce the same image positions. The configuration derived through our initial estimation and subsequent optimisation represents one possible mass distribution that matches the observed image positions. A parameter-by-parameter comparison across the full sample is therefore not an appropriate test of reliability. Published parameters for these systems are not unique: they depend on the treatment of the group environment and the number of mass components adopted. WFI2033-4723 illustrates this, as it lies in a galaxy group and is modelled with several explicit mass components plus group shear \citep{rusu_2020} rather than a single SIE+shear configuration. Furthermore, $b^\prime$ is not equal to the Einstein radius for non-zero ellipticity, so a direct numerical comparison is meaningful only for the near-circular systems.

While additional observational constraints -- e.g. the ratio of image fluxes or time delays -- could refine these configurations and provide a more accurate representation of the lens mass distribution, this level of refinement is not necessary for X-ray astrometry. The mass models we calculate for these lensed systems are sufficient for performing milliarcsecond X-ray astrometry, although they are, in general, not expected to be detailed enough for other applications, including time-delay cosmography. This is because X-ray astrometry depends only on precise image position predictions which are determined by the gradient of the potential, whereas time-delay cosmography depends on the overall shape and depth of the lens potential, which require more tightly constrained and physically realistic models. What matters for the caustic-based X-ray astrometry that motivates this work is the accuracy with which the positions of the optical lensed images (used to constrain the lens model) are reproduced. The case of B1422+231 demonstrates this: smooth SIE+shear models reproduce its image positions readily, and fail only in the image fluxes, which arise from substructure \citep{brada__2002} and which our method does not use as inputs.

\subsubsection{External shear -- ellipticity degeneracy}\label{subsec:ellipdegen}
The simulated training data does not include external shear. However, we introduce this parameter during the optimisation process to better capture the diversity of possible image configurations. Given the (widely known) degeneracy between ellipticity and external shear, allowing small variations in the shear parameter increases the flexibility of the SIE model and enables it to fit a wider range of lensing configurations. Therefore, most resulting models are ellipticity dominated and cannot have large external shear. However, our approach effectively reproduces most of the simulated observed image configurations and allows accurate reproduction of the position of the four lensed images, as required for milliarcsecond X-ray astrometry. Nevertheless, the approach of not considering external shear in the simulated data may not be representative of real lensing galaxies. This was suggested by \cite{luhtaru_what_2021}, who found that, out of a sample of 39 observed quads, 15 were shear-dominated and 11 were ellipticity-dominated. 

In future, for our methodology, it would be beneficial to simulate a training set that includes external shear and to attempt to infer an initial estimate for this parameter. 

\subsection{Optimising Mass Models}\label{subsec:MassModelOpt}
We attempted to optimise 200 simulated gravitationally lensed systems that the neural networks had never seen, each with randomly generated ellipticity angles. Of these 200 simulated lenses, 151 converged to a lens configuration with a chi-squared value of less than 8. The optimisation results are outlined in Table \ref{tab:optimisationResults}. With current and prospective future X-ray telescopes, we can reach a maximum resolution of image position measurements of 0\farcs5. Out of the 200 simulated quads, 179 had image separations greater than 0\farcs5, and 75 per cent of these converged to high-fidelity lens configurations able to reproduce the original image positions to milliarcsecond accuracy. 

Errors on the position estimate of the centre-of-mass of the lens did not have a significant effect on the number of models optimised. 79 per cent of models with an error on the mass centre of $< 0\farcs5$ converged, whereas 72 per cent of models with an error on the mass centre of $> 0\farcs5$ converged.

The RMS errors on $b^\prime$ and $\epsilon$ increase from 0\farcs149 $\rightarrow $ 0\farcs206 and from 0.145 $\rightarrow $ 0.260 respectively for lens configurations that did not converge, a markedly larger separation than for the centre-of-mass error above. The predicted $b^\prime$ and $\epsilon$, together with the centre-of-mass position and ellipticity angle estimated from the image positions, provide the initial values from which the optimisation proceeds. Convergence therefore appears to depend more strongly on the accuracy of the neural network predictions than on that of the initial centre-of-mass estimate. The poorer these initial estimates, the less likely a high-fidelity lens configuration can be recovered within a reasonable time-scale.

\subsection{Parameter Predictions}\label{subsubsec:ParameterPredictions}
For a few select lens configurations across the mass parameter and ellipticity range, we compared the predicted mass parameter and ellipticity against their known values. The neural network prediction errors varied as a function of both the source position and the lens configuration. Across all lens configurations, the prediction of parameters near the cusps had a significant error, even though this area was heavily sampled in the training data (Sec. \ref{subsubsec:gridding}). This is likely due to a greater difference in resulting image positions when a source near the cusps is varied slightly, compared to the same variation of position of a source near the lens centre \citep[see Figs 3 and 4 of][and corresponding discussions in the paper]{spingola_milliarcsecond_2022}. 

The prediction error on simulated lensed systems with added ellipticity rotation was significantly higher in comparison to lensed systems with zero ellipticity rotation. The RMS errors on the mass parameter and ellipticity predictions were 0\farcs1481 and 0.0299 respectively for non-rotated systems (Table \ref{tab:NeuralNetworkResults}); 0\farcs165 and 0.180 respectively for randomly rotated systems (Table \ref{tab:optimisationResults}). There is a noteworthy change in the ellipticity prediction for rotated systems ($\approx 6\times$ increase), which introduces variations in lens orientation to more closely resemble what is observed in actual lensing systems. If the rotation angle were calculated to greater accuracy, this error could be significantly reduced and the number of converged lens configurations would increase. This could be improved by locating the real mass centre of the lensing galaxy, which is typically revealed by observations in the radio or optical bands, added as an output to the neural network predictions, or found by following the methods of \cite{schechter_even_2019} as described in Sec. \ref{subsec:lenscentre}.

\subsubsection{Mass Parameter}\label{subsubsec:DiscussMass}
Figures \ref{fig:sourceprederrorM25} and \ref{fig:sourceprederrorM55} plot the prediction error of the mass parameter for $10^6$ randomly distributed sources in the top right quarter of the inner caustic for two lens configurations with no ellipticity rotation or shear. As expected, the prediction error increases when a source is closer to the cusp (Sec. \ref{subsubsec:gridding}). Sources close to the edge of the caustic result in lower prediction errors, likely due to a high density of training data around the caustics.

Across all lens configurations, a significant variation in the prediction accuracy for both $b^\prime$ and $\epsilon$ was observed throughout the source plane enclosed by the caustic. While only one quadrant of the caustic is shown in Figs.~\ref{fig:sourceprederrorM25} and \ref{fig:sourceprederrorM55}, we note that the sharp transition in prediction accuracy is present symmetrically across all four quadrants of the caustic interior, with the position and magnitude of this transition depending on the specific mass parameter and ellipticity of the lens configuration. This phenomenon arises from the image labelling method employed (see Sec. \ref{subsec:networkfeatures}), where the image with the highest flux is designated as image A. The remaining images are labelled B through D in either a clockwise or anti-clockwise sequence, determined by the proximity of the next closest image to image A. This labelling strategy ensures consistency under rotations and translations, maintaining uniformity in the features used to train the neural network. However, at the labelling discontinuity line within the caustic interior, the relative brightness ordering of the two most highly magnified images changes, producing two geometrically distinct input states for the same lens configuration. Since the network cannot distinguish between these two input states from the image coordinates alone, the conflicting mappings learned during training produce the sharp discontinuity in prediction accuracy seen in Figs.~\ref{fig:sourceprederrorM25} and \ref{fig:sourceprederrorM55}. The position and severity of this discontinuity vary with the lens configuration, as the relative magnifications of the lensed images, and therefore the point at which the brightness ordering changes, depend on both $b^{\prime}$ and $\epsilon$. The most direct remedy would be to adopt a purely geometric image labelling scheme that does not rely on flux ordering, which would eliminate the labelling discontinuity entirely by ensuring a consistent input representation regardless of the relative image brightnesses. In principle, this effect could also be mitigated in future work by augmenting the network input features to explicitly encode the flux ordering of the images, or by training separate networks for each possible labelling state. Alternatively, since the discontinuity line divides the caustic interior into two distinct regions, a classification step could be introduced prior to regression to determine which labelling state applies for a given source position.

The distributions of predicted mass parameters for each mass are plotted along the y-axis in Fig. \ref{fig:massViolin}. In each violin, the shaded region represents the probability density of the predictions at each value, with wider regions indicating a higher concentration of predicted values; the central box and whiskers denote the interquartile range and median respectively, and the red lines mark the 2.5th and 97.5th percentiles. Note that the width of each violin is fixed and therefore the widths cannot be compared between different mass parameters. Our simple neural network architecture predicts well across the range of mass parameters ($b^\prime=0\farcs1$--5\farcs5). However, the prediction distribution modes for mass parameters below 1\farcs5 were slightly above the identity line, whereas for mass parameters above 2\farcs0, the prediction modes fell below the identity line. A method to improve the prediction accuracy would be to apply a linear correction to the mass parameter predictions to reduce their distance from the identity line. This could be applied separately for mass parameter predictions above 2\farcs0 and below 1\farcs5.

Across all mass parameters, 95.5 per cent of the neural networks' predictions were within 0\farcs3 of their true value. The predictions for $b^\prime = 0\farcs1$ to $b^\prime = 0\farcs3$ were all incorrect and would require retraining the model to accurately predict. This behaviour is consistent with edge effects in the training data, where fewer effective examples are available and the network is required to extrapolate. While extending the training range and retraining the network could in principle alleviate these effects, this was not pursued here, as mass parameters approaching zero are not relevant to observed systems in the context of strong gravitational lensing. Nevertheless, the maximum error in this regime was $0\farcs31$, which remains well within the range that can be corrected during the optimisation process. As the lensed image separation scales with mass parameter, small mass parameters span a much smaller region of the parameter space than larger mass parameters; consequently, the model performs less reliably for the lowest values of $b^\prime$.
 
Overall, given the purpose of estimating initial mass model parameters, the predictions from our neural networks are acceptable and could be further refined when optimising the mass model.

\subsubsection{Ellipticity}\label{subsubsec:DiscussEllip}
Fig. \ref{fig:ellipViolin} plots the distribution of ellipticity predictions for each ellipticity value across all lens configurations. The mean ellipticity error across all models was $\pm$0.0217 and the mode of each distribution is within $\pm$0.05 of the identity line. However, a small number of predictions for every ellipticity have an error in excess of $\pm$0.2. The identity line falls within the $25^{\text{th}}$ and $75^{\text{th}}$ percentile of each distribution for ellipticities from 0.1 to 0.55. For ellipticities at the lower and upper limits of the training domain ($\epsilon \lesssim 0.1$ and $\epsilon \gtrsim 0.55$), the predicted values slightly over- or under-estimate the ellipticity; however, the mode of the predictions remains within $\pm$0.05 of the true values. We note that the slight bias observed at the lowest and highest ellipticities is consistent with a systematic divergence of the network predictions near the limits of the training domain, where fewer effective examples are available. While such effects can in principle be mitigated by extending the training range, this was not pursued here, as the ellipticity values in our training set ($\epsilon=0.05$--0.60; Sec.~\ref{subsec:simulations}), were chosen so as to cover the ellipticities of the majority of elliptical galaxies \citep{chen_ellipticities_2016}. Instead, these edge effects could be reduced in future work by increasing the sampling density near the edges of the physically motivated parameter space, or by applying simple bias corrections to the network predictions.

Overall, the predictions for ellipticity are acceptable for a first estimate of the parameters assuming no external shear. They will be further refined during the optimisation process using the \texttt{lensmodel} software, requiring them to be accurate enough to constrain the parameter space that the lens configurations could cover, such that the optimisation can occur in a reasonable time frame without getting stuck in local minima. This `accuracy' is arbitrary, and any improvement on the prediction errors will decrease the time taken for models to optimise as well as increase the number of models that can converge to a low chi-squared value. However, with 75 per cent of simulated unseen systems converging to a chi-squared of less than 8 with an ellipticity prediction RMS of 0.145, we consider the predictions to be acceptable as initial parameters. 

\subsubsection{Microlensing Effects}\label{subsubsec:DiscussMicro} 
Quasar microlensing is typically caused by stars in the lensing galaxy crossing in front of one of the lensed images, for which its observed flux is magnified. Since the Einstein radii of microlenses are very small compared to the distance between light paths of different images, changes in flux and in position at the microarcsecond level occur independently for each observed image. While the mass parameter and ellipticity predictions are invariant under microlensing, the initial labelling of `image A' is determined by the image with the highest absolute flux. If one or more images are microlensed, the image subsequently labelled `image A' may no longer correspond to the image with the largest flux magnification. Therefore, in such a scenario, the relationships that the neural network uses to make predictions would be incorrect and result in an erroneous parameter estimation. This potential image mislabelling could be resolved by analysing the image configuration, as the two most magnified images typically exhibit the smallest separation, and therefore the prediction can be done twice using each image as `image A'. A more robust solution, which we leave to future work, would be to embed this disambiguation directly into the pipeline, for instance via the flux-independent labelling scheme discussed in Sec. \ref{subsubsec:DiscussMass}, which would remove the flux dependence entirely. We refer the reader to \citet{vernardos_microlensing_rev_2024} for a review of quasar microlensing. 

\subsection{Propagated Errors}\label{subsec:DiscussPropErrors}
All the calculations within the neural network architecture are exact and have no associated errors. Therefore, the predictions are values with no known errors. However, the inputs to the network are measured values, such as the positions of the lensed images, and as a result they have uncertainty in their measurements. These errors are not propagated or taken into account during the parameter prediction process. If associated errors on the mass parameter, ellipticity and ellipticity angle can be calculated, then the parameter space to cover during the optimisation process can be significantly constrained.

The simulated data the neural networks learn from also have no simulated errors, as all image positions and fluxes are calculated precisely. As the neural networks do not train on data that have measurement errors or uncertainties, their performance on real data may be affected.

While we believe error quantification is very important, particularly for machine learning and parameter inference, this paper aims to use limited initial data to produce initial estimates of mass model parameters to speed up the model creation and optimisation process. We recognise that there are methods to infer parameter errors for fully connected neural networks, but they fall outside the scope of this work.

\section{Conclusions}\label{sec:Conclusions}
X-ray-to-optical astrometry of quadruply lensed AGN provides a fruitful avenue for investigating the AGN multiplicity and sub-structure at sub-kpc distances at early cosmic times \citep[$z \lesssim 4$; see discussions in sections 6.3 and 6.4 of][]{siskreynes_2026}. Here, we have demonstrated a machine learning-based approach to infer suitable initial parameter values for SIE mass models, aimed at facilitating forward modelling of strongly lensed systems using the lens equation. To the best of our knowledge, this is the first work to explicitly demonstrate that lens modelling can be accelerated by exploiting the geometric properties of lensed image positions alone, without relying on full imaging data. We note that while image fluxes are used in the ordering and labelling of the lensed images prior to network input, they are not used as direct inputs to the neural network. Unlike full Bayesian inference frameworks such as \textsc{GIGA-Lens} \citep{Baltasar2026} -- which employ GPU-accelerated multi-stage MCMC sampling to fully characterise the posterior distributions of lens parameters -- our approach provides rapid initial lens model estimates using a simple machine learning method that requires no GPU acceleration, making it well suited for large-sample applications.

Previous work in the machine learning lensing literature has largely focused on using lensing images as input to convolutional neural networks (CNNs) for tasks such as candidate identification or direct parameter inference. We have presented a fully connected neural network approach that, by ingesting only a very limited amount of information can predict the mass and ellipticity of the lens with enough accuracy to significantly reduce the computation time for optimisation with \texttt{lensmodel} routines in \textsc{gravlens} software. Specifically, we are able to predict initial parameters for a lens mass model to sufficient accuracy that a high-fidelity mass model can be refined within a few minutes for 75 per cent of simulated SIE cases. High-fidelity mass models were also refined for seven real gravitationally lensed systems, with a mean optimisation time of 142 seconds per model. We have demonstrated that the use of simple neural networks to characterise lensing systems can significantly speed up the inference computations. This will facilitate a timely investigation of the over 200,000 prospective gravitational lensed systems to be observed in the next few years with upcoming surveys, in particular in cases for which lens equation optimisation for an arbitrary potential benefits from a data-informed initial estimation of the parameters.

We also highlight some limitations of our approach. While the networks consistently overestimate mass parameters of 0\farcs1 or 0\farcs2, 95 per cent of mass parameters predicted for these values are within 0\farcs130 and 0\farcs104 of the true value respectively. Ellipticity predictions are significantly affected by the error estimation of the rotation angle of the lens ellipticity, which resulted in an increase in the ellipticity prediction RMS from $0.030$ to $0.180$ (an absolute increase of $0.150$, corresponding to a factor of ${\sim}6\times$), evaluated across 200 randomly rotated lens configurations. Improving this angle estimation would result in a significantly greater accuracy of ellipticity predictions. 

There are other improvements beyond the scope of this investigation that could increase the number of lensed systems that would quickly converge to a high-fidelity mass model. These include adding external shear to the neural network training set. Shear predictions could allow for more image position variability and models that represent shear-dominated lensing galaxies. The models in the training data could also have randomly set ellipticity angles, which the networks could try to predict. Our simulations also provide the raw materials for an Approximate Bayesian Computation (ABC) approach that would allow posterior retrieval on the lens mass model parameters.

\section*{Acknowledgements}
\label{Acknowledgements}
This work was supported by NASA Contract NAS8-03060 to the \axaf\ X-ray Center, NASA ADAP grant 80NSSC24K0617, and the SAO/University of Southampton Astronomy Masters Program. RMG was supported by AstroAI and AstroMind. The authors thank the referee and Cecilia Garraffo for useful comments.

\section*{Data Availability}
The positions of the seven lensed quasars mentioned in Section \ref{sec:results} are available on the \gaia~Archive and contained within the public \gaia~Data Release 3. The full code pipeline to reproduce the simulations, neural network training and parameter predictions, along with the trained network models and optimisation code, are publicly available on GitHub at \url{https://github.com/aostridge/Grav-Lens-ML}. Other reduced data products used in this work may be shared on reasonable request to the corresponding author.

\bibliographystyle{mnras}
\bibliography{references}

\clearpage
\appendix
\renewcommand{\theequation}{A\arabic{equation}}

\section{Machine Learning}
\label{sec:appendixmachinelearning}
Here, we define some standard terms and concepts from machine learning that are relevant to the methods discussed in the main text.

Machine learning focuses on building algorithms and models that can learn patterns from data, without a human explicitly programming them. One such machine learning algorithm is the neural network, which is loosely based on how the biological neurons in the human brain work. The key hyper-parameters of a neural network (values that control how the network learns) are the number of neurons, the number of neuron layers, activation functions, and the selected optimisation method. Each neuron is a node in the network that takes inputs from the previous layer, applies linear operations and a non-linear activation function to them, and relays the result to neurons in the following layers. Layers refer to groups of neurons found at the same depth in the network. Each neuron in a given layer is connected to each neuron in the following layer by a scalar weight, which determines the strength of the connections (a multiplicative factor on each of the inputs) -- allowing for individual adjustments of the relationships between parameters. Every neuron also applies a bias to its output, which offsets the output slightly and allows for more variability in the network. 

The other main component of neural networks is the optimisation method, which minimises the loss function by updating the weights and biases of the network. The loss function quantifies how poorly the model is performing at any given moment, for example the mean squared error of the predictions compared to true values. One of the most popular optimisation methods is Stochastic Gradient Descent\citep[SGD;][]{tran-dinh_gradient_2022}, and we use a variant of this method in this project. 

While machine learning can infer relationships from data, it generally requires a large amount of input data to train from. To get accurate results, the inputs have to be representative and unbiased as the model can only learn from the data it is shown. However, once trained, the model can predict values from given inputs very quickly, making it well suited to estimating values for large datasets.

\subsection{Loss Function -- Stochastic Gradient Descent} \label{appendixLossFunction}
The loss is a function of the variable parameters of the network (weights and biases). The most direct method to minimise the loss function is to calculate its gradient across the network with respect to all weights for all input training data. Using this, the minima of the function can be found, which would specify the network parameters that minimise the loss function. However, this process demands significant computational resources and time. Instead, SGD estimates the gradient of the loss function for each example of training data \citep{ruder_overview_2017}. By iteratively feeding the network training examples, recalculating the gradients of the loss function, and updating the weights and biases, the network can reach a minimised solution. This may not always result in the most direct path to the minimum, as by using a single set of inputs instead of the entire input dataset, SGD can only calculate an approximation of the true loss function gradient. As a compromise, we use a variant of SGD that uses small batches of input training data when approximating the loss function. This reduces the total number of updates to the weights and biases, decreasing the computational demand, while approximating the loss function more accurately. 

There is a learning parameter ($\eta$) that determines the step size along the gradient of the network for each iteration. A higher learning rate can lead to quicker convergence; however, it increases the chance of overshooting the minimal loss function. Low learning rates can result in the network getting stuck in local minima and increase run time. This trade-off can lead to using an adaptive learning rate that is large at the start of training and reduces as convergence is approached.

The process of updating the weights and biases in the network to minimise the loss function is called back-propagation. This is calculated for each node by finding the partial derivative of the loss function with respect to each weight. The partial derivatives are calculated from the output layer backwards towards the input layer, terming it back-propagation.

Propagating the inputs from the first layer through the network, and applying the weights, biases and node operations, yields the output values. These can be compared to the true values based upon the specific inputs to calculate a loss function. The network then uses back-propagation to edit the weights and bias of the network to attempt to reduce the loss function. One of these loops is defined as an `epoch', and this is repeated for many epochs until the loss function is minimised.

We can utilise a `dropout layer' to help prevent overfitting -- where a network learns to reproduce the learning examples well but loses the ability to generalise to unseen examples. Dropout layers randomly set inputs to 0 with a set frequency \textit{f} during each epoch, and all inputs that are not set to 0 get scaled by $1/(1 - f)$ so that the sum of inputs is the same (where \textit{f} is a float between 0 and 1).

\subsection{Activation Function -- ReLU} \label{appendixReLU}
Activation functions are non-linear functions applied at each node before `passing on' the output to the next layer. These allow the network to approximate more complex functions.

In this work, we specifically employ the rectified linear unit (ReLU) activation function, which is widely used for its simplicity and effectiveness. ReLU is applied to the outputs of each node and outputs 0 for all negative inputs, passing positive inputs unchanged. It is one of the most popular activation functions in machine learning when predicting continuous parameters, due to the ease of deriving its gradient (as it is a straight line). This is useful for back-propagation where the derivative of the loss function is calculated (which depends on the activation function's derivative). It can also give different activation strengths based upon the alignment of the inputs $\Bar{x}$ and weights $\Bar{w}$, since the output equals their dot product plus the bias $b$ of the node.
\begin{equation}
    \centering
    \text{output} = \Bar{w}\cdot\Bar{x} + b \label{eq:outputnode}
\end{equation}

\noindent The derivative of ReLU is not defined at zero, so either it is set to 0 (which we use for the network), or a soft-ReLU function is used that is continuous at $x=0$. We use the ReLU function as we are trying to solve a regression problem, where we are predicting a continuous value. Also, setting negative inputs to zero effectively reduces the number of activated neurons on each propagation through the network, and can reduce the computational load and help prevent overfitting.

\section{Computational Setup}
\label{appendix:computational_setup}
All computations were performed on a Dell Inspiron 7501 running Ubuntu Linux (23.04--24.04), with an Intel Core i7-10750H CPU, 16 GB RAM, and a 2 TB SSD. All code was executed on CPU only, without CUDA or GPU acceleration; reported runtimes correspond to this reference machine.
\label{lastpage}
\end{document}